# Topological surface states in dense solid hydrogen

Ivan I. Naumov[1] and Russell J. Hemley[1,2]

[1]*Geophysical Laboratory, Carnegie Institution of Washington, Washington DC 20015, USA*

[2]*Lawrence Livermore National Laboratory, Livermore CA 94550, USA*

It has recently been established that two-dimensional massless graphene-like systems and three-dimensional line-node topological semimetals comprise a special class of centrosymmetric materials where edge/surface states of topological nature inevitably appear in some **k**-regions. We show that the phases of solid hydrogen produced at megabar pressures can be line-node topological semimetals. In such phases, the material exhibits topological surface states and therefore can be a poor bulk metal but good surface conductor. The results may help to explain discrepant high-pressure experimental data reported for dense hydrogen as well as provide predictions for future measurements, including possible surface superconductivity in hydrogen. Related topological behavior may be expected in high-pressure phases formed in other elements including simple metals.

In his seminal 1939 paper Shockley showed how surface states (SSs) appear in one-dimensional (1D) centrosymmetric crystal as the lattice parameter *a* decreases [1]. At some critical point $a_c$, the valence and conduction and valence bands cross, leaving two surface states within the bulk energy gap; at the $a_c$ the latter closes and then immediately opens up becoming "inverted". Much later, Zak realized that Shockley-type SSs have in fact a topological character [2]. He demonstrated that when the surface coincides with the symmetry centers, the existence or absence of the corresponding SSs is decided by the topology of the band below the gap. These results are found to be relevant to understanding graphene and other carbon-based structures. Because of the connections between graphene and structures of dense hydrogen, it is of interest to consider the existence of SSs in the latter. We show that compressed hydrogen can exhibit occupied topological SSs in its high-pressure semimetallic states. This surprising property of the material significantly affects the electrically conducting properties of the material and should be taken into account in the interpretation of experimental results. Indeed, the findings reported here may help to resolve controversies surrounding previously published data on this fundamental system.

Of special interest in the general theory of SSs is the case when the surface is taken between the atoms, which is the situation that Shockley discussed in his pioneering work. In this case, speaking in more contemporary terms, the existence of SSs is related to the sum of all Zak's phases *Z* below the gap [3,4]. The SSs exist when the total phase is π and does not exist when it is 0. From the topological point of view, Shockley's critical parameter $a_c$ is nothing but the point of topological transition at which the Zak's phase discontinuously changes from 0 to π. Though Shockley and Zak considered only a simple 1D model, their results can be extended to systems of higher dimensions. As established recently, criteria for the existence of SSs are still applicable in centrosymmetric zero-gap semiconductors where the conduction and valence bands touch each other at points (2D *massless graphene-like* systems [5-7]) or along lines (3D *line-node topological semimetals* [8,9]). In such 2D and 3D materials, the existence of SSs with a particular momentum $\mathbf{k}_\parallel$ is controlled by the value of Zak's phase $Z(\mathbf{k}_\parallel)$ obtained by the integration across the Brillouin zone (BZ) perpendicular to the edge/surface [5-9]. As in the 1D case, the SSs exist if $Z(\mathbf{k}_\parallel)$ is π (inverted band gaps) and do not if it is 0. For this, however, the phase *Z* becomes dependent on a new parameter $\mathbf{k}_\parallel$, which can be critical similar to the lattice parameter *a* in the 1D case. The possible values of $Z(\mathbf{k}_\parallel)$, 0 or $\pi$, are completely defined by the positions of the band-contact points or lines representing singularities in the electrons dispersion. In the 2D case, for example, the Zak's phase $Z(\mathbf{k}_\parallel)$ undergoes a jump whenever the integration path crosses the line connecting two band-contact points [5-7]. Thus, in graphene with zigzag edges, the SSs states exist within a finite $\mathbf{k}_\parallel$-interval corresponding to the projection of the Dirac points $\mathbf{K}$ and $\mathbf{K}'$ on the $\mathbf{k}_\parallel$ axis; within this interval $Z(\mathbf{k}_\parallel) = \pi$ [5-7].

The line-node topological semimetals (LNSs) can be viewed as 3D analogs of graphene. Instead of two separated Dirac points, they have an infinite number of effective Dirac points merged together − band-contact lines lying at the Fermi level exactly (nodal lines) [8-10]. In LNSs, the SSs appear within the whole $\mathbf{k}_\parallel$-area where $Z(\mathbf{k}_\parallel) = \pi$; this area is limited by the projection of the nodal line (loop) onto the surface plane of interest [8-10]. The SSs are completely dispersionless and therefore characterized by an infinite density of states (DOS) at the Fermi level [8-10]. This unphysical DOS obtains from an idealization of the LNSs, namely, from the assumption that band-contact line lies at the Fermi level [8]. In real materials, the probability that band-contact lines to coincide with the Fermi level is vanishingly small because the mechanisms that stabilize such lines differ from those that force them to have a constant (Fermi) energy [8]. Many systems, however, can be considered as "approximate" LNSs where the contact lines have some dispersion and so do the SSs. An example is rhombohedral graphite with ABC stacking where the nodal line is replaced by a chain of connected electron and hole pockets [10,11]. Though SSs in "approximate" LNSs acquire some dispersion, they still retain their topological nature and their existence can be predicted by the bulk Zak's phase [10,11]. Remarkably, the bulk Zak's phase is capable of capturing not only the difference between the surfaces of distinct orientations, but also the difference between the distinct terminations (if any) for a given orientation [6,7,9]. Such a nontrivial correspondence between the Zak's phases and surface terminations appears because the surface is made by cutting a solid *between the primitive unit cells* [5-7] (we will refer to this as a *cutting rule*). This rule says that once the surface orientation and termination are specified they automatically specify the bulk primitive unit cell.

A consideration of simple elements is useful for experimental realization of these phenomena. Notably, the majority of the known "approximate" LNSs are represented by carbon materials, including rhombohedral graphite mentioned above. In the latter, the topological SS's states form on the (0001) surface, if the hexagonal setting is used [10,11]. Such states also form on the zigzag lateral surfaces of usual hexagonal graphite with AB stacking [12,13], and on the (110) plane in a model carbon allotrope with a symmetry *Cmcm* ($D_{2h}^{17}$) [14]. A cleaner example may be found in the high-pressure phases of solid hydrogen. Recent experimental [15-22] and theoretical [23-30] studies reveal that structures of compressed hydrogen above 200 GPa can be viewed as layered structures consisting of distorted graphene-like sheets. Dense molecular hydrogen phases resemble the well-known carbon phases such as graphene and graphite not only *structurally*, but also *electronically* because the 1*s* electrons in a hydrogenic honeycomb lattice and 2*p_z* electrons in real graphene lead to topologically similar band structures [31-33]. By analogy with the carbon systems, it is reasonable to expect that compressed hydrogen is also capable to exhibit occupied topological SSs once it reaches a semimetallic state. Should such SSs exist, they would drastically change the conductive properties of the system; this fact must be taken into account in experiments aimed to detect the onset of metallic state via direct electrical resistance measurements. Indeed, some anomalous data may point to these issues.

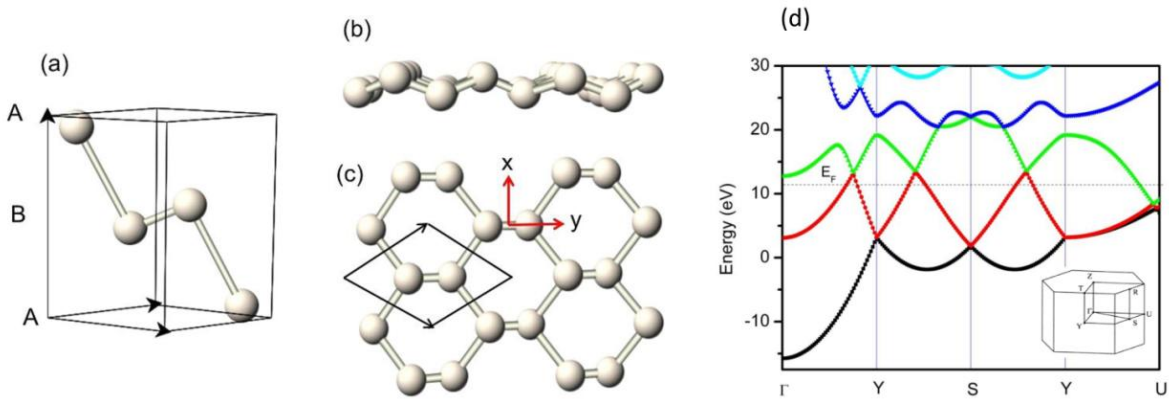

**Figure 1**. Crystal and band structures of the *Cmca*-4 phase. (a) primitive unit cell, (b) side and (c) top of a single layer. (d) Segments that involve the Y- point. The conduction and valence band intersections do not coincide with the Fermi level, indicating that the Fermi surface consists of small electron and hole pockets.

To illustrate our point, we choose the molecular hydrogen structure *Cmca*-4 as an example. This structure has predicted to stability between 250 and 500 GPa [23-27], and has been used to interpret experimental data. Geometrically, the structure is close to the hexagonal graphite − its primitive unit consists of four atoms and its layers are arranged in an ABAB sequence (Fig. 1). The layers A and B are identical but shifted with respect to each other by some vector lying in the *xy* plane. Each layer can be viewed as a buckled graphene layer consisting of $H_2$ molecules tilted relative to the *xy* plane θ by ~ $30^0$. The structure exhibits inversion symmetry, which in combination with time-reversal symmetry guarantees the quantization of the Zak's phase [7].

The calculated band structure of *Cmca*-4 hydrogen at 300 GPa is shown in Fig. 1d. One can see that it is close to that in a zero band-gap semiconductor. The conduction and valence bands intersect each other in a linear (Dirac-like) fashion for all the segments starting from the Y-point. Since the Dirac-like points must lie on band contact lines normal to the corresponding segments [35], one can expect that they all belong to *one* band-crossing loop lying in the *xy* plane and encircling the point Y. Such a loop should induce the SSs localized on a (001) surface. There are two possible terminations associated with this surface orientation. The first one breaks the long (weak) interlayer chemical bonds, whereas the second breaks the short (strong) intralayer bonds. According to the cutting rule, these two terminations correspond to two different primitive unit cells shown in yellow rectangles in Fig. 2 (panels *a* and *b*). Hereafter, we call these terminations/unit cells "type *a*" and "type *b*". Since we treat the surfaces as the boundary planes of the films repeated periodically in space at equal vacuum gaps, such films should be built from the complete bulk primitive unit cells as building blocks. Their thickness therefore can be specified by the number of bulk primitive cells stacked along the *z* axis (*n*). It is easy to see that an *n*-unit-cell thick film organized from type *a* unit cells contains *n* molecular layers A and *n* molecular layers B (see Figs. 1 and 2). At the same time an *n*-unit-cell thick film organized from the type *b* unit cells contains *n*-1 molecular layers A, *n* molecular layers B (both inside the film), and additionally two atomic surface layers.

To check that the band-crossing loop and the corresponding SSs do exist, we calculated the phases $Z(\mathbf{k}_\parallel)$ across the BZ in the *z* direction for both choices of the unit cell (Figs. 2 c, d). We treated our system as if it were an ideal LNSs, *i.e.* we assigned the number of "occupied" bands at each $\mathbf{k}_\parallel$ to be 2 (see the Supplemental Material [34]). For both cases the Zak's phase is quantized; *i.e.* it takes only values 0 and π. For the type *a* case, the Zak's phase is 0 almost everywhere in the surface BZ except relatively small areas around the $\overline{Y}$ points (panel *c*). In going from the type *a* to type *b* case, all the $Z(\mathbf{k}_\parallel)$ values shift by π , so that all the 0 phases become π and vice versa. Due to intimate relation between the $Z(\mathbf{k}_\parallel)$ values and the SSs, these results suggest that in the type *a* case the SSs will cover a relatively small ellipse-like area around the $\overline{Y}$ point, with the major radii almost reaching the $\overline{U}$ points, whereas in the type *b* case− a significant part of the surface BZ centered at $\overline{\Gamma}$.

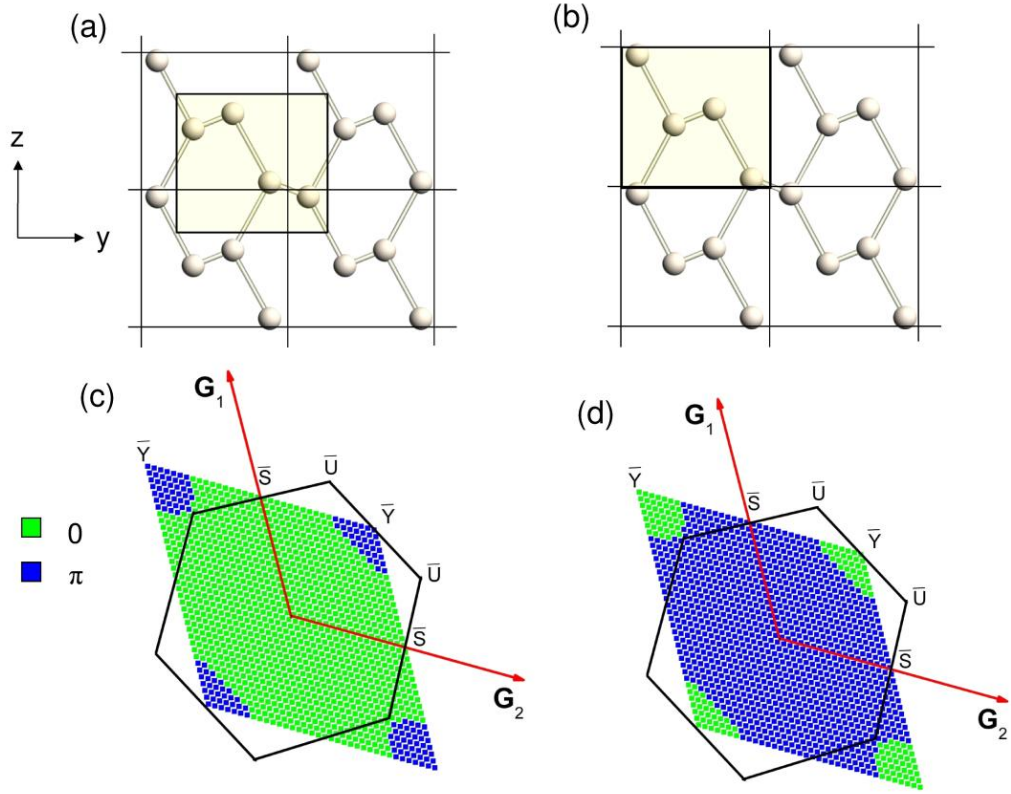

**Figure 2**. (a,b) Two choices of the primitive unit cell corresponding to two different surface terminations parallel to the *xy* plane, shown by yellow rectangles. For simplicity, we show the unit cells projected on the plane that contains all the 4 atoms shown in Fig. 1a. (c and d) Zak's phase Z as a function of $\mathbf{k}_\parallel$ for the unit cells (a) and (b), respectively. The $\mathbf{k}_\parallel$-points are given on a 36x36 grid associated with the two in-plane reciprocal vectors, $\mathbf{G}_1$ and $\mathbf{G}_2$. The k-point labels correspond to those presented in Fig. 1d. The phases 0, π are shown in green and blue, respectively.

The direct calculations of surface electronic structure for 4-unit-cell thick films of H show that this is indeed the case (Fig. 3); the calculations for thicker films lead only to slightly different quantitative results. The surface bands obtained can be identified indicated by of thick black curves that cross the Fermi level. For type *a*, they appear around the $\overline{Y}$ point, in agreement with Fig. 2*c*. In contrast, for type *b* they are realized everywhere in the surface BZ, but not in the vicinity of the $\bar{Y}$ point, again in accordance with the Zak's phase calculations. We thus see that the stronger the broken bonds, the larger is the **k**-area covered by the corresponding SSs; this allows one to interpret the SSs as dangling bond surface states. Notably, upon moving away from their "natural habitats", the SSs gradually *turn* into bulk states, as shown by the thick red curves in Fig. 3. This reflects their topological nature and intimate relationship to the bulk electronic structure. Note that the wiggles in the dispersion curves around the $\bar{Y}$ and $\bar{S}$ points stem from the finite thickness of the films. As the thickness increases, the wiggles become smaller in scale because the two surface states at each **k** approach to each other in energy.

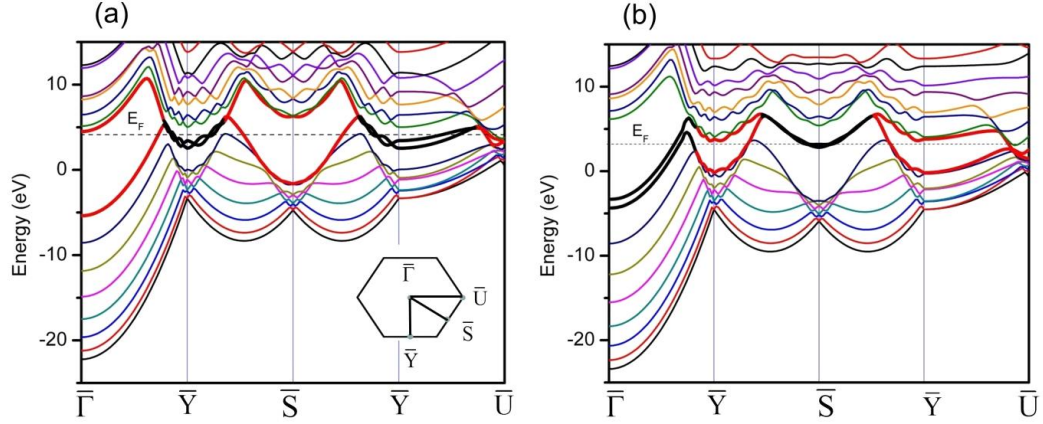

**Figure 3**. Two-dimensional energy bands of hydrogen in a 4-unit-cell thick *Cmca-4* structure. (a) Terminations that break long (weak) bonds on both sides of the film and (b) terminations that break short (strong) bonds. The surface bands are indicated by of thick black curves that cross the Fermi level.

As seen from Fig. 3, the SSs always appear in pairs, almost in the middle of the corresponding band gaps. One surface band falls out of the allowed bulk *valence* band and the other from the allowed *conduction* bulk band. Due to the presence of inversion symmetry, they are either symmetric or anti-symmetric with respect to the operation **r**→−**r.** They do not coincide in energy because the opposite sides of the film still "feel" each other. However, they become exactly degenerate as the film becomes infinitely thick. It is straightforward to prove using Shockley's arguments [1] that the SSs must be *metallic*. Indeed, in a *Cmca*-4 (001) hydrogen film there is *one more* band in the lowest, valence part of the spectrum relative to the similar effective bulk system. Therefore, this additional (surface) band must be partially occupied with electrons.

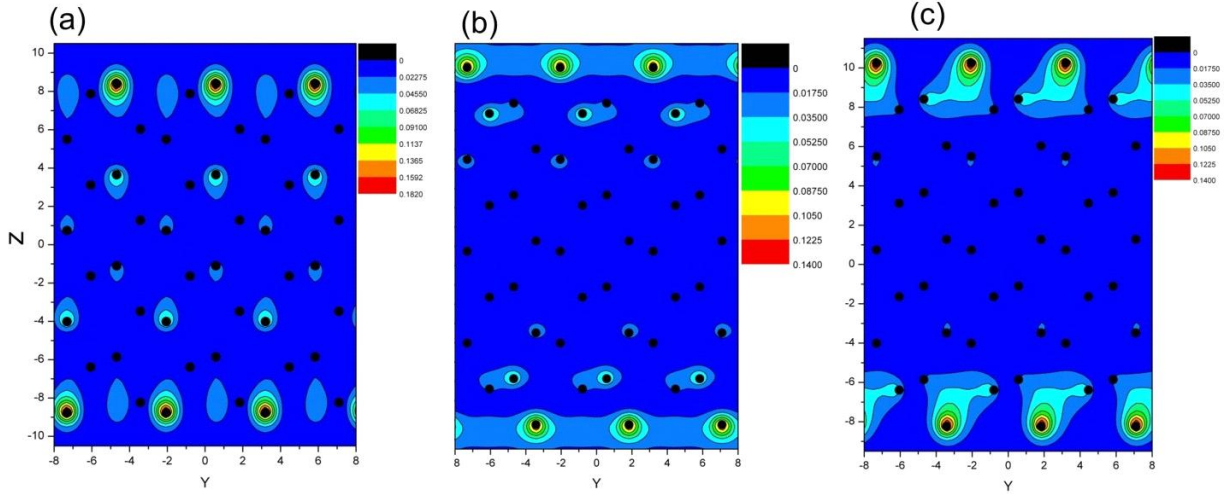

**Figure 4**. The charge density of the lower (out of 2) surface state at different high-symmetry points in the surface BZ and different types of terminations, for 4-unit-cell-thick films in the $x=0$ plane. (a) at the $\bar{Y}$ point for the *type a* termination. (b and c) at the $\bar{\Gamma}$ and $\bar{S}$ points, respectively, for the type *b* termination. Full black circles indicate the atomic positions. The surface atoms correspond to the top and bottom layers along the *z* direction. Note that the $x=0$ plane coincides with that shown in Figs. 2 a,b. The charge density of the upper (in energy) surface state is very similar to that for the lower state, and this difference disappears when the film thickness increases.

The charge distributions of the lower SS at different high-symmetry points in the $x=0$ plane are given in Fig. 4. One can see that in the case of the type *a* termination, the density is mainly localized in the two upper atomic layers, whereas for type *b* termination it is in the first and third surface layers (Fig. 4 a,b). Further, it easy to see that the charge on the opposite surfaces is located on the *different non-overlapping sublattices.* All sublattices corresponding to one surface can be obtained from the sublattices corresponding to the opposite surface by the inversion symmetry (**r**→−**r**). This situation is similar to that in graphene ribbons where the amplitude of the edge states from one side is located only on one, say A, sublattice, whereas on the other side−on B sublattice. The metallization of the surface for type *b* termination seems natural: in this case the molecular (strong) bonds on the surface are broken. The situation, however, is not as trivial for case *a* termination, which preserves the "$H_2$" molecular structure of the surface layer. Even in this case the electronic structure of the layer is *metallic* in character and the wave function is localized essentially on only one of the two H atoms (Fig. 4a).

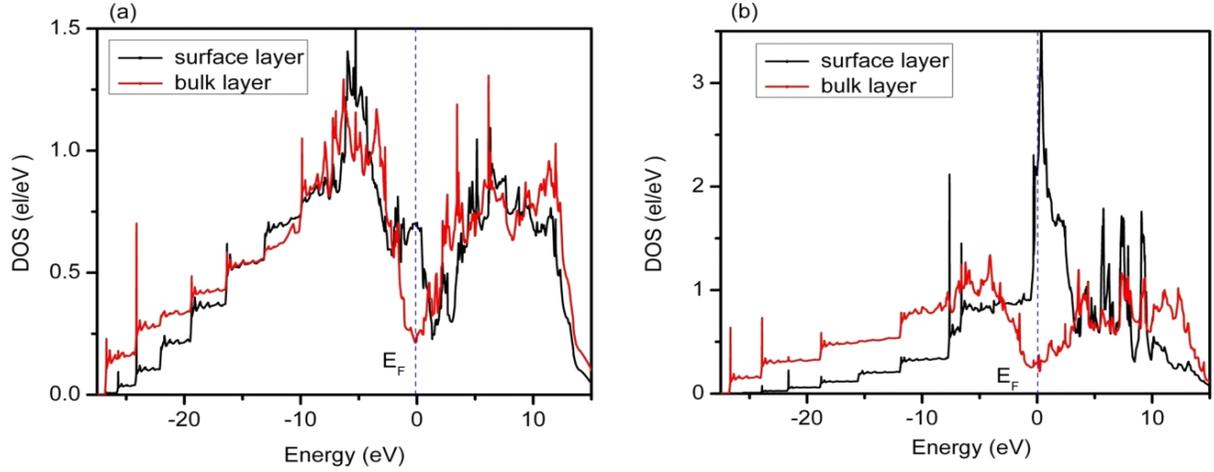

**Figure 5**. The local density of states (DOS) in hydrogen sphere of a radius 0.74 Bohr in a 4-unit-cell thick film of *Cmca*-4 hydrogen. (a) Film with type *a,* and (b) type *b* terminations. The black curves correspond to the most top surface atoms and the red curves- to the atoms lying in the central layer.

As seen from Fig. 5, the local DOS at the Fermi level significantly increases upon formation of the (001) surfaces−by a factor of 3 and 10 for type *a* and *b* terminations, respectively. The local DOS for the atoms belonging to the central layer is typical for bulk *Cmca*-4 hydrogen, where the $E_F$ is in the minimum of DOS. The increase in this quantity as one approaches the surface of the type *a* is explained by the surface states around the $\overline{Y}$ point (Fig. 3 a). The sharp peak in the DOS just at $E_F$ in the case of *b* termination is associated with the flat surface band near the $\overline{S}$ point (Fig. 3b).

Looking at the Zak's phase distribution $Z(\mathbf{k}_\parallel)$ (Figs. 2*c* and 2*d*), one can imagine the situation when the band-crossing loop shrinks to the Y point and then completely disappears. This means a semimetal-to-insulator transition when the band gap opens at Y. It is clear that in such an insulating state the Zak's phase will be zero for type *a* termination and $\pi$ for type *b*, in both cases *for all* $\mathbf{k}_\parallel$ in the *xy* plane. Correspondingly, the SSs will no longer exist for the first termination, but will cover the whole surface BZ for the second termination. Such a situation is realized, in other high-pressure hydrogen structures that are both theoretically predicted and experimentally observed, e.g., *C*2/*m* [36,37] and *Pca*2$_1$ [23,24,38,39]. In these two structures, like *Cmca*-4, the molecules H$_2$ are also tilted relative to the *xy* plane. The (001) surfaces thus can have different terminations− "molecular " and "atomic" −depending on whether they break weak (intermolecular) or strong (molecular) bonds. As numerical calculations show, all the molecular surfaces are insulating whereas atomic−metallic. Thus, the phases *C*2/*m*, and *Pca*2$_1$ with atomic (001) surfaces behave similar to topological insulators−they are insulating in the bulk but necessarily have metallic surface states.

In this context, it is interesting to proceed from *C*2/*m* to another similar candidate phase (*Cmc*2$_1$ [36]), which also has tilted H$_2$ molecules but does not have a centre of symmetry and therefore develops a spontaneous polarization. As for *C*2/*m*, the (001) surfaces in *Cmc*2$_1$ are insulating or metallic, depending on termination. In contrast to *C*2/*m*, however, now *Z*(**k**$_∥$) cannot be 0 or $\pi$ for all **k**$_∥$, due to lack of the inversion symmetry. Instead, calculations show that their averaged over **k**$_∥$ values (Berry's phases [40]) become −0.0013 and 0.9987$\pi$, respectively; the slight deviation from 0 and $\pi$ indicates a very weak polarization in *Cmc*2$_1$. The coexistence of ferroelectricity and topological SSs makes the *Cmc*2$_1$ a unique material along with the recently discovered CsPbI$_3$ [41].

The present predictions should be considered in the interpretation of previously reported experimental data, as well as new results being obtained, for dense solid hydrogen. Early Raman experiments reported anomalous optical spectra localized on the hydrogen-diamond interface [42]. Differences in the degree of optical properties of hydrogen samples from different experiments at pressures above 250 GPa have been reported [43,44]. In addition, there have been questions about quantitative matching of the increase in reported electrical conductivity [15] with the observed changes optical properties based on changes in bulk properties [32]. These results may point to role pressure-induced changes in surface states, rather than bulk properties as has been typically assumed. Moreover, topological SSs similar to those discussed here for hydrogen can also form in some simple *sp* metals. For example, the pronounced Shockley states found earlier in Be on its (0001) surface around the $\bar{\Gamma}$ point [45] could have been also predicted from Zak's phase calculations. Due to their topological nature, these states are relatively flat; according to Ref. [46] the local DOS at the Fermi is four times larger than in the bulk. It has been speculated that such a situation can trigger an unusual form of surface superconductivity with a high *T*$_c$ [46]. By analogy, one may expect such surface superconductivity in compressed hydrogen.

# Topological surface states in dense solid hydrogen

Ivan I. Naumov[1] and Russell J. Hemley[1,2]

[1]*Geophysical Laboratory, Carnegie Institution of Washington, Washington DC 20015, USA*

[2]*Lawrence Livermore National Laboratory, Livermore CA 94550, USA*

We used a norm-conserving pseudopotential as implemented in the ABINIT package [1]. The norm-conserving pseudopotential was generated with OPIUM codes using the Perdew-Burke-Ernzerhof GGA functional. The cutoff radius $r_c$ was chosen to be 0.5; it is less than ½ of the shortest interatomic distance in the structure *Cmca*-4 at 300 GPa. A cutoff energy of 100 Ry was used for the plane-wave expansion of the valence and conduction bands wave functions. A 40×40×40 Monkhorst-Pack **k**-point grid has been used in the case bulk band structure calculations. The hydrogen surfaces were simulated by a system of parallel films separated by vacuum gaps. The value of a vacuum gap were chosen to be equal to eight interlayer distances to ensure negligible wave function overlap between the replica films. A 40×40×1 Monkhorst-Pack **k**-point grid [2] was used in the surface calculations where the third dimension corresponds to the surface normal.

To calculate $Z(\mathbf{k}_\parallel)$, we modified the standard ABINIT code in the that deals with the electronic polarization. This part handles usual ferroelectrics (dielectrics) where the phase $Z$ must be continuous function of $\mathbf{k}_\parallel$. Accordingly, the code prevents the function $Z(\mathbf{k}_\parallel)$ from unphysical jumps including those from one "polarization branch" to another [3]. We removed this restriction because in our case of LNSs such jumps are allowed. Another subtlety is connected with the fact that in "approximate" LNSs, in contrast to ferroelectrics, the number of occupied bands at different $\mathbf{k}_\parallel$ points can be different due to existence of bits of Fermi surface. But since we are interested in the topological properties of the bands, we treated our system as if it were an ideal LNSs, *i.e.* we assigned the number of "occupied" bands at each $\mathbf{k}_\parallel$ to be 2 and exploited the usual formula for $Z(\mathbf{k}_\parallel)$

$$Z(\mathbf{k}_\parallel) = \frac{-i}{2\pi} \sum_{n=1}^{M} \int_0^{\mathbf{G}_\perp} \langle u_{n\mathbf{k}} | \partial_{\mathbf{k}_\perp} | u_{n\mathbf{k}} \rangle d\mathbf{k}_\perp \qquad (1)$$

where $\mathbf{G}_\perp$ is the shortest reciprocal lattice vector perpendicular to the surface, $\partial_{k_\perp}$ is the partial derivative along the $\mathbf{G}_\perp$, $u_{n\mathbf{k}}$ is the periodical part of the Bloch wave function *n*, and the summation is over all the occupied bands. Generally, the surface of interest may not necessarily be perpendicular to some reciprocal lattice vector $\mathbf{G}_\perp$. For the example of *Cmca*-4, this is the case because we consider only the (001) surface.